\documentclass{PoS}
\usepackage[backend=biber,sorting=none]{biblatex}

\title{Universality and template synthesis of cosmic ray air shower radio emission}

\ShortTitle{EAS radio template synthesis}

\author{David Butler$^{a}$, \speaker{Tim Huege}$^{bc}$, Ralph Engel$^{ab}$ and Olaf Scholten$^{cd}$\\
		\llap{$ˆ{a}$} Institut f{\"u}r Experimentelle
Teilchenphysik (ETP), Karlsruhe Institute of Technology (KIT)\\
		PO Box 3640, 76021 Karlsruhe, Germany\\
		\llap{$ˆb$} Institut f{\"u}r Kernphysik (IKP), Karlsruhe Institute of Technology (KIT)\\
		PO Box 3640, 76021 Karlsruhe, Germany\\
		\llap{$ˆc$} Astrophysical Institute, Vrije Universiteit Brussel\\
		Pleinlaan 2, 1050 Brussels, Belgium\\
		\llap{$ˆd$} Center for Advanced Radiation Technology (KVI), University of Groningen\\
		Zernikelaan 25, NL-9747 AA Groningen, The Netherlands\\
        E-mail: \email{david.butler@kit.edu}, \email{tim.huege@kit.edu}, \email{ralph.engel@kit.edu}, \email{scholten@kvi.nl}}

\abstract{Accurate prediction of the radio emission from cosmic ray air showers relies on computationally demanding Monte Carlo simulations such as CoREAS. We aim to expedite this process via a semi-analytical synthesis model while maintaining high accuracy by using simulated radio pulses as templates. We present our key concept for template processing focusing on the development of the particle cascade and its empirical effect on the locally produced radio signal. In this context the universality of the radio emission from small sections of an air shower also becomes important where most previous studies focus on integral quantities observable at far distances.

}

\FullConference{36th International Cosmic Ray Conference -ICRC2019-\\
		July 24th - August 1st, 2019\\
		Madison, WI, U.S.A.}

\begin{document}

\section{Introduction}

On one hand our current understanding of radio emission from cosmic ray air showers requires a full microphysics Monte Carlo (MC) simulation to accurately interpret measurements. On the other many analysis techniques work very well using incredibly simple approximations such as assuming a point source located at the shower maximum $X_\mathrm{max}$ while many aspects of the underlying particle cascade can be reparametrised to appear highly universal \cite{Lafebre:2009en}. In this work we describe our efforts to find such a universal parametrisation of the emitted radio signal with the primary intent of using it in a fast and accurate alternative to radio MC codes such as CoREAS. Such a model is particularly valuable to experiments where many antennas are illuminated by a single event as the computation time of CoREAS scales with the number of antennas which becomes prohibitive for LOFAR and SKA.


Common theories of radio emission from cosmic ray air showers assume two macroscopic mechanisms: induction of transverse currents by charge separation in the Earth's magnetic field, and a time-varying negative charge excess as electrons are swept from air molecules while the positive ions are left behind \cite{2016PhR...620....1H}. Due to their different natures the geomagnetic emission is polarised along the $\vec v \times \vec B$ direction while the polarisation of charge excess emission is aligned radially. Even though the individual emission profiles should be radially symmetric to first order their coherent sum is not, thus in order to study both separately we evaluate the radio signal along the $\vec v \times \vec v \times \vec B$ axis where the two mechanisms produce orthogonal polarisations.

In the lateral direction the signal is known to be continuous and smooth for small differences in distance allowing linear interpolation of amplitude and phase spectra \cite{Ewa:diplom}. Combined with the radial symmetry of the amplitude and known polarisation patterns this allows reconstructing the full lateral plane from a single line of discrete sample positions. Such procedures are already implemented in existing analytical models such as MGMR \cite{Scholten:2007ky} and forms the basis of most lateral distribution models in analysis. The difficulty lies in predicting the radio signal at any one sample position. At large viewing angles emission from different evolution stages of the cascade arrives in chronological order forming a broad pulse with low peak amplitude. Very close to the shower core the highly relativistic cascade overtakes the radio signal because the refractive index in air does not perfectly equal unity leading to an inverted timing hierarchy \cite{2011PhRvL.107f1101D}. For many geometries the transition between these two regimes produces a ``Cherenkov ring'' where signals from almost all parts of the air shower arrive simultaneously leading to a short but very high pulse. We capture these complex interactions by directly reprocessing detailed MC simulations as template showers. 

Unless otherwise specified all plots and stated parameter values in this work refer to showers induced by vertical proton primaries of $10^{17}$eV. All shown radio signals were filtered to 0-500 MHz using a rectangular bandpass filter while ensuring our time resolution satisfies the Nyquist sampling criterion for these frequencies. While some coherent radio emission exists at higher frequencies most experiments operate at 30-80 MHz with the LOFAR high-band stations up to 200 MHz \cite{2015APh....65...11N} and the prototype SKA-low antenna \cite{2015ICRC...34..309H} up to 350 MHz being the primary experimental incentives for considering higher frequencies.
\section{Longitudinal Slicing}
\begin{figure}
\begin{center}
\includegraphics[width=0.4\textwidth]{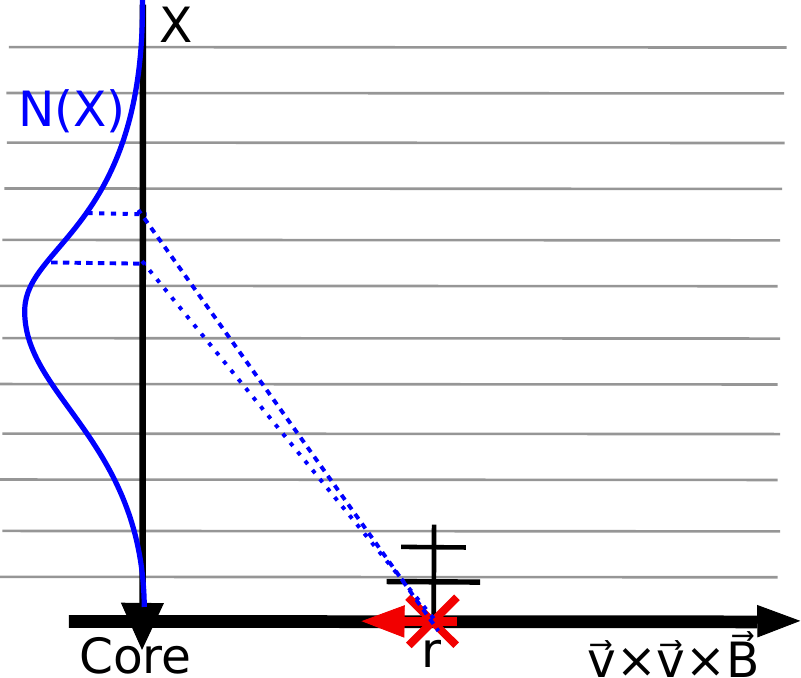}
\includegraphics[width=0.45\textwidth]{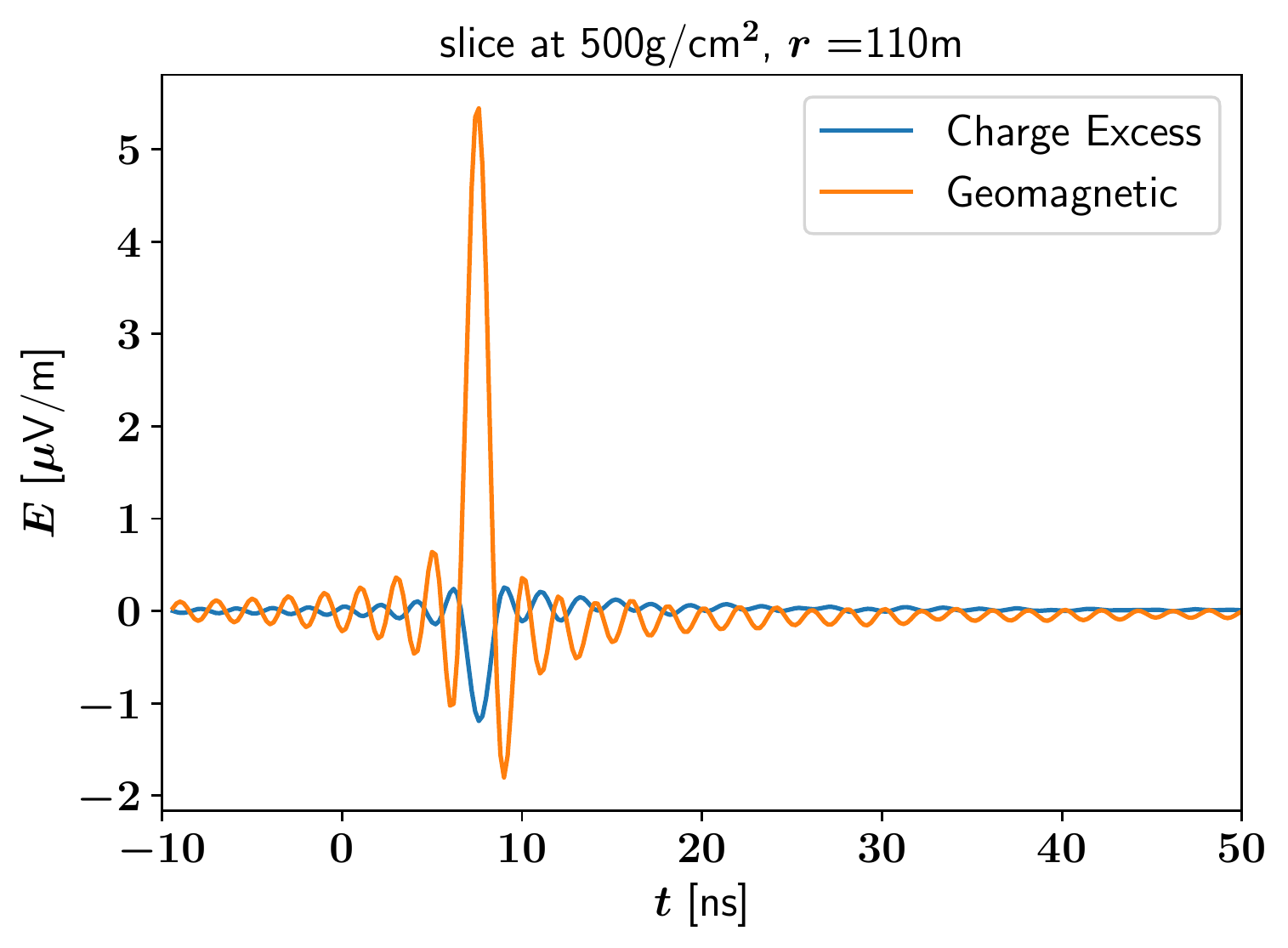}
\includegraphics[width=0.45\textwidth]{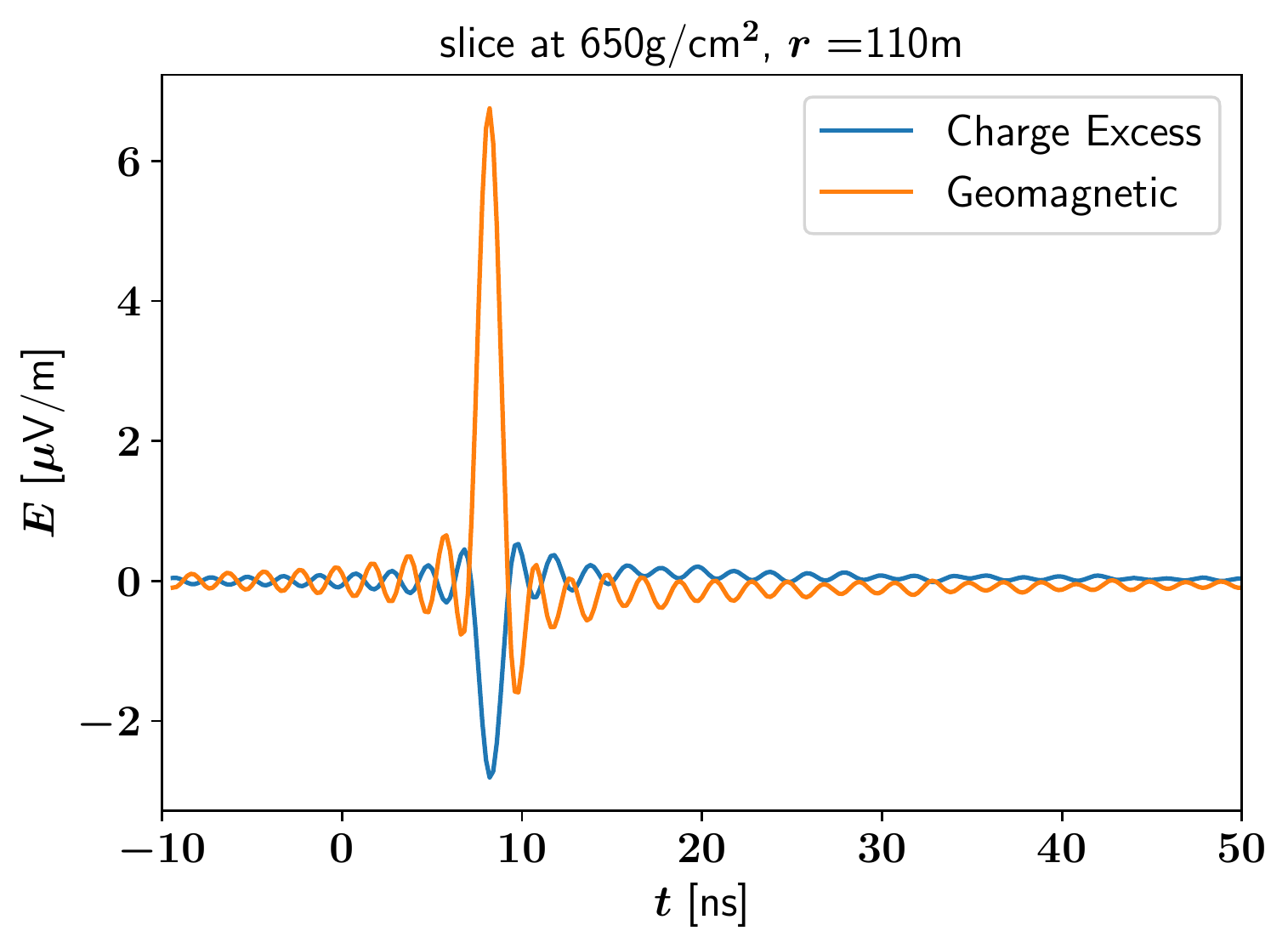}
\includegraphics[width=0.45\textwidth]{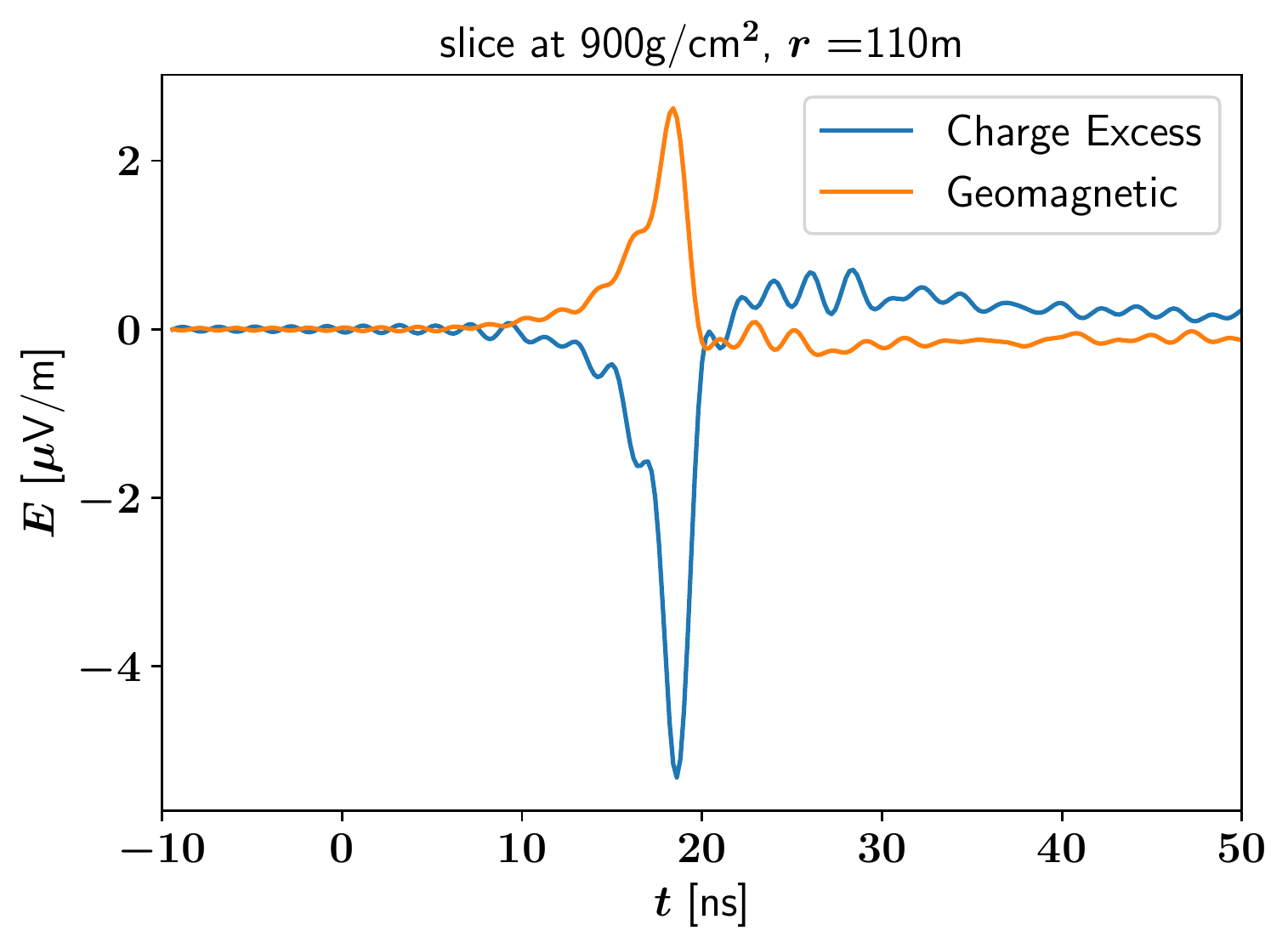}
\caption{Top left: schematic showing slicing of radio signal by atmospheric depth of contributing particles. Others: individual slice radio time series for three slices before, close to and after the shower maximum all for the same antenna positioned on the Cherenkov ring. Note both the viewing angle \textit{and} the locally applicable Cherenkov angle vary from slice to slice therefore these time series are not expected to be equal.}
\label{fig:Xslice}
\end{center}
\end{figure}

An obvious starting point for template construction is the evolution of the particle cascade. We choose to define slices in slant depth $X$ which for vertical showers is a direct (but non-linear) mapping of longitudinal distance, elapsed time or height above ground. Particles in the cascade are assigned to exactly one slice as if it were one macroscopic source while we still evaluate the radio emission from all particles within each slice using the full endpoint formalism in CoREAS, a schematic and sample slice radio signals can be seen in Fig. \ref{fig:Xslice}. The simulation is configured so that the direct sum of all individual slice radio signals yields the physically observable electric field at a given position $\vec r$:
\begin{equation}
{\vec E}_\mathrm{phys} (\vec r, t) = \sum_X {\vec E}_\mathrm{slice}(X, \vec r, t)
\end{equation}

The electric field amplitude is assumed to scale linearly with the inverse of the propagation distance. Deviations from this simple approximation can first be observed at about 3.5 km distance to the ``source'' at which point the internal structure of the particle cloud becomes visible.

\section{Simple Rescaling Synthesis}

In accordance with existing studies \cite{2016PhR...620....1H} and consistent with CoREAS we only count electrons and positrons as sources of radio signals. The particle number $N$ is sampled directly for each slice but is in very good agreement with the well-known Gaisser-Hillas curve \cite{1977ICRC....8..353G} fully defined by 3-5 scalar parameters. Assuming a constant ambient density within the slice the coherent electric field is expected to be directly proportional to the particle number. Thus we can rescale the signal from each slice individually to map a template shower (Temp) onto a target real shower (Real):
\begin{equation}
{\vec E}^\mathrm{Synth}_\mathrm{slice}(X,\vec r,t) = \frac{N^\mathrm{Real}(X)}{N^\mathrm{Temp}(X)} \cdot {\vec E}^\mathrm{Temp}_\mathrm{slice}(X,\vec r,t)
\end{equation}
which can then be summed to produce a synthesised observable signal (Synth)
\begin{equation}
{\vec E}^\mathrm{Synth}(X,\vec r,t) = \sum_X\frac{N^\mathrm{Real}(X)}{N^\mathrm{Temp}(X)} \cdot {\vec E}^\mathrm{Temp}_\mathrm{slice}(X,\vec r,t).
\end{equation}
\begin{figure}
\begin{center}
\includegraphics[width=0.4\textwidth]{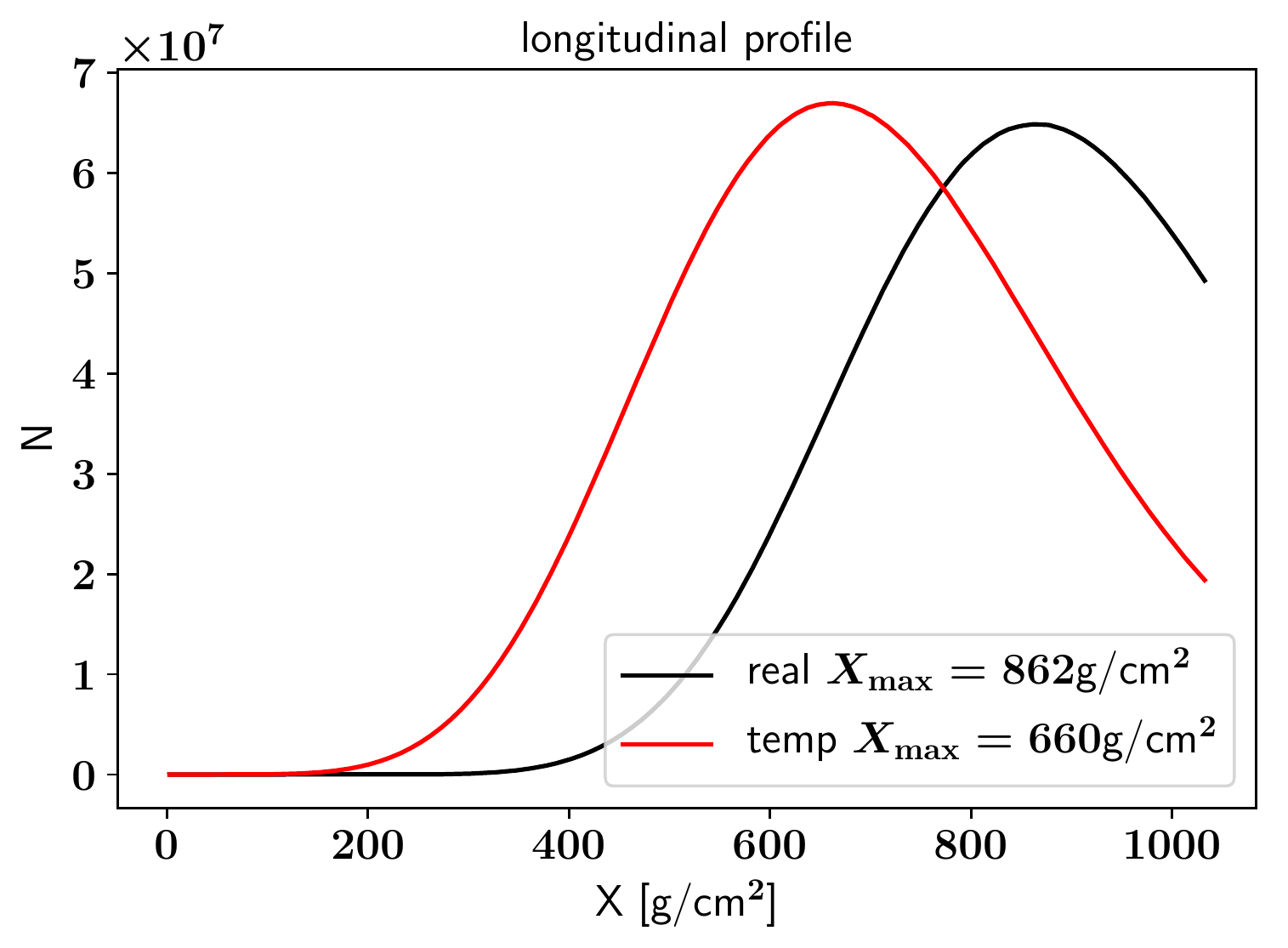}
\includegraphics[width=0.45\textwidth]{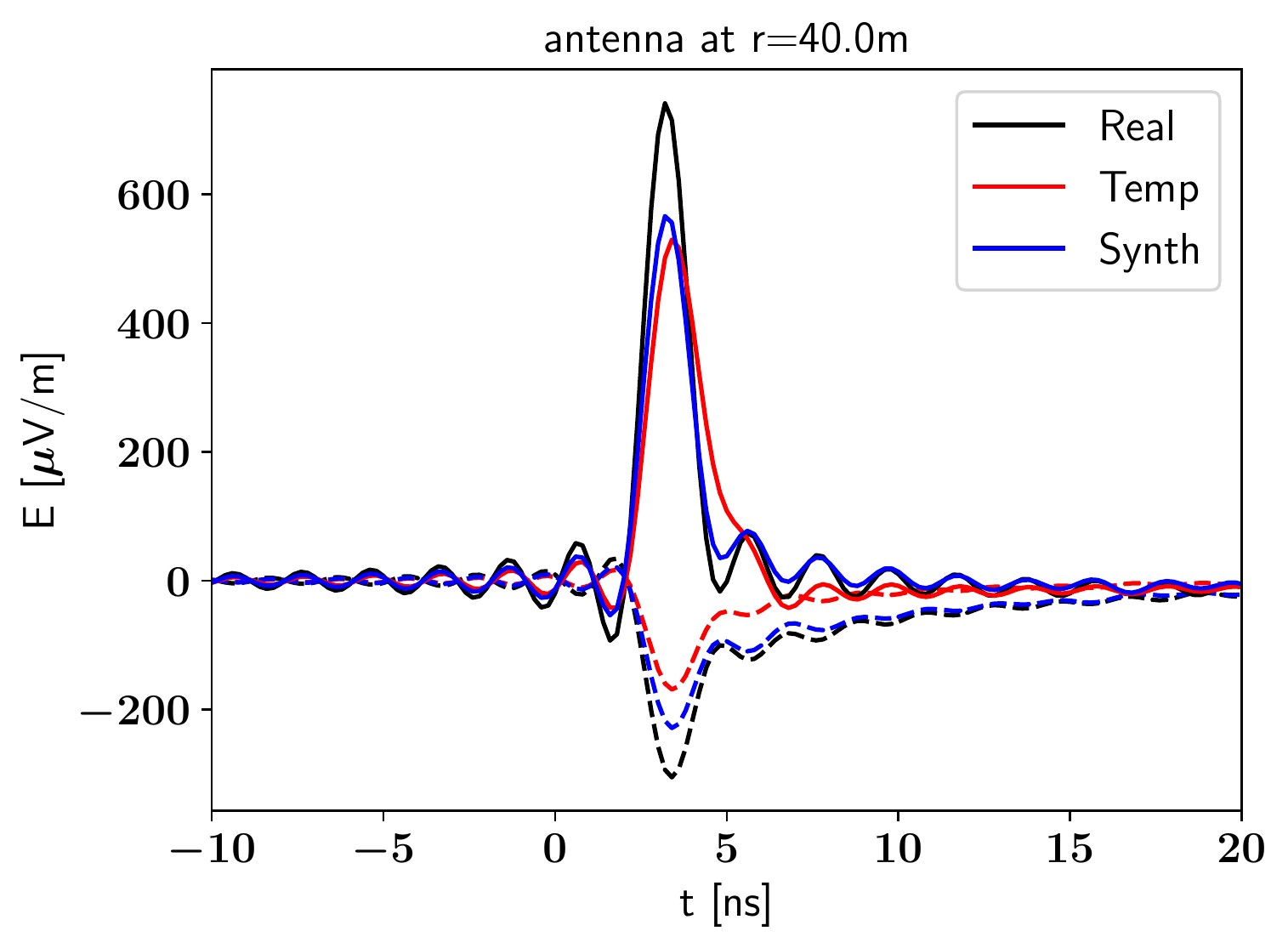}
\includegraphics[width=0.45\textwidth]{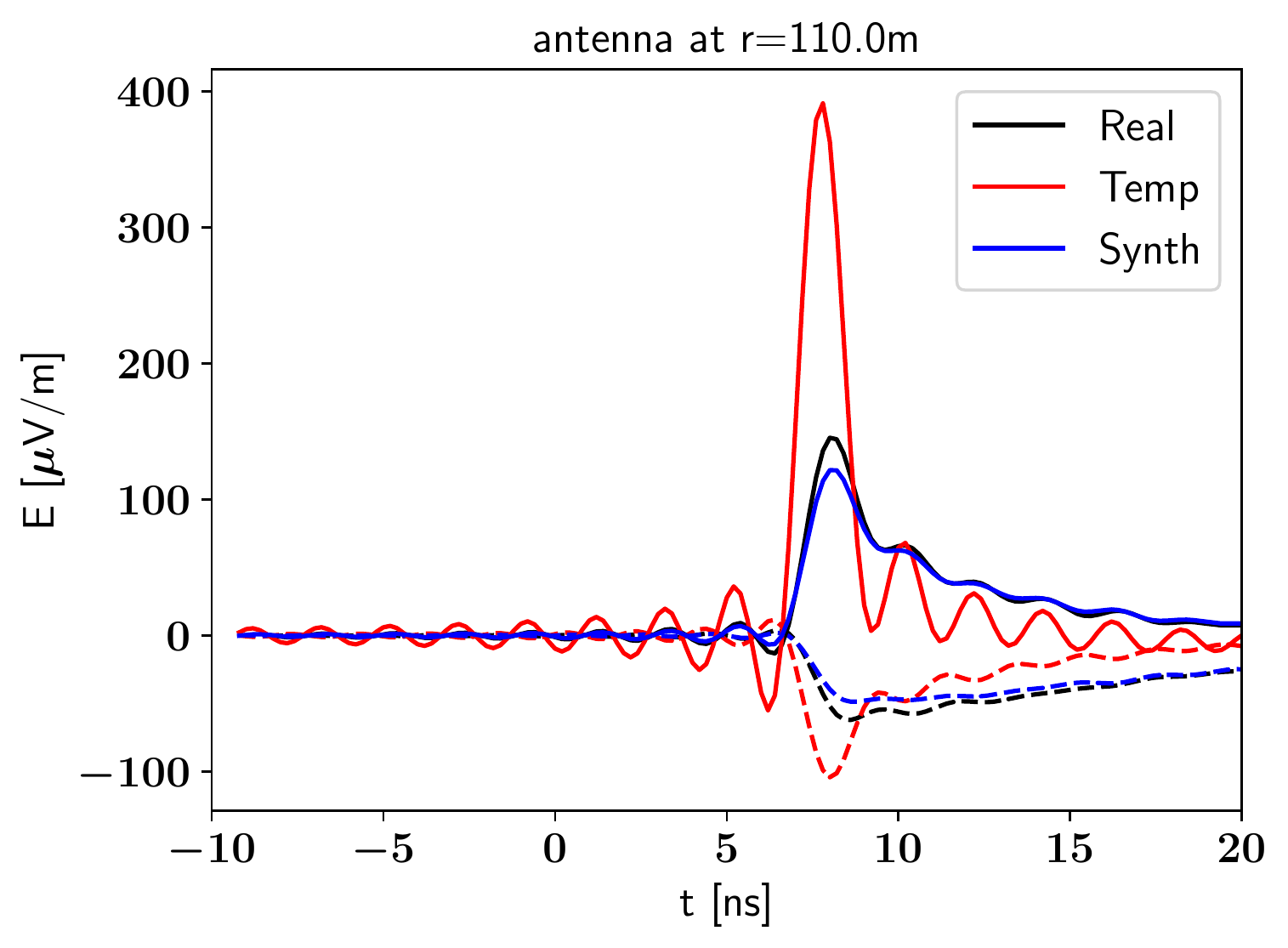}
\includegraphics[width=0.45\textwidth]{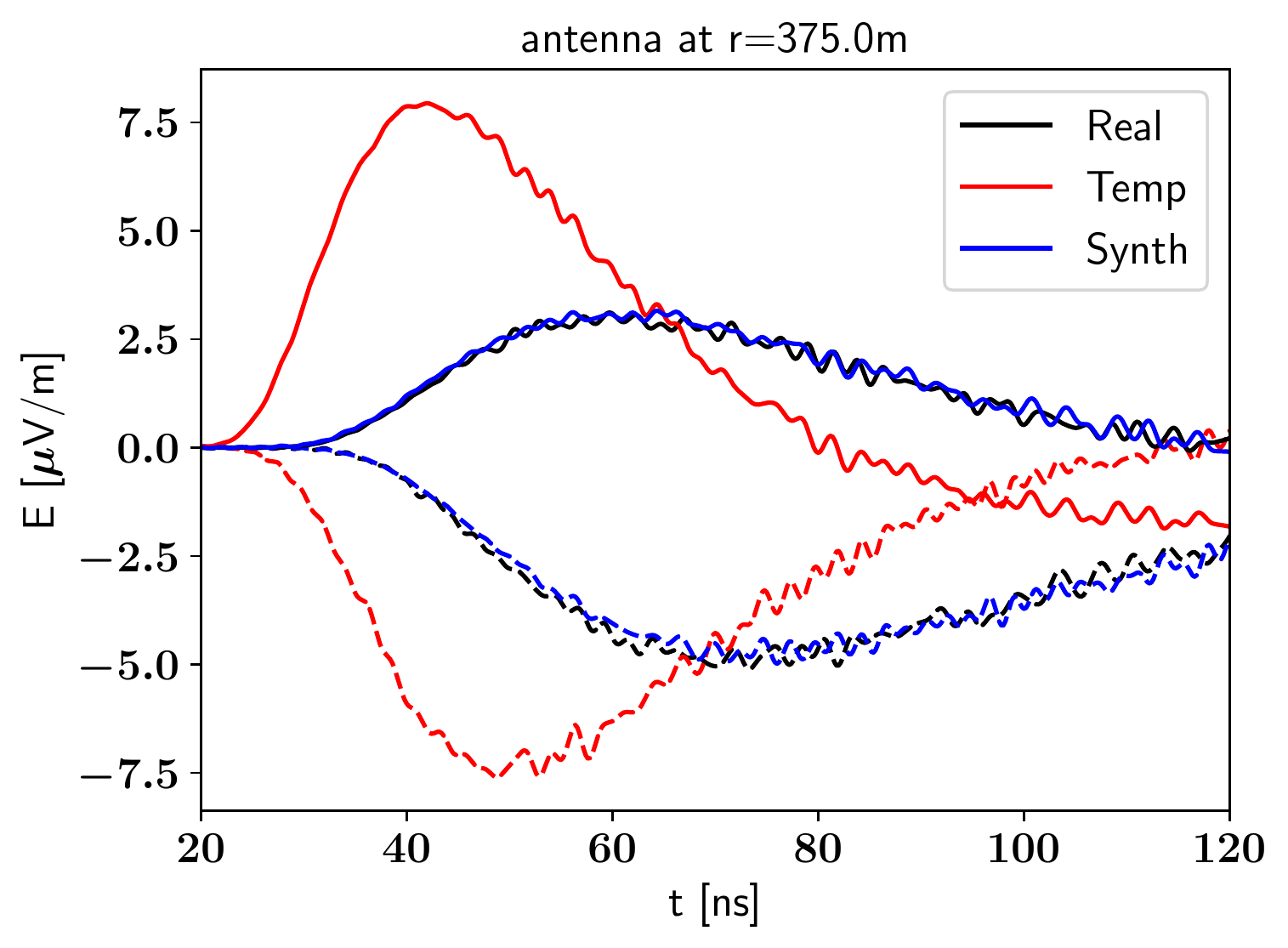}
\caption{Simple rescaling synthesis. The longitudinal profiles of the template and target are shown in the top left. The radio signals and synthesis result at three different lateral distances inside (top right), on (bottom left) and outside (bottom right) the Cherenkov ring can be seen for both polarisations corresponding to Geomagnetic (full stroke) and Charge Excess (dashed) emission. }
\label{fig:ssynth}
\end{center}
\end{figure}

As shown in Fig. \ref{fig:ssynth} this works encouragingly well already. Of particular note is the near-perfect synthesis for large lateral distances far beyond the Cherenkov ring suggesting a highly universal behaviour. There appear to be two main contributing factors to this observation: first of all the Cherenkov beaming of the emission from each individual slice leads to only earlier stages of the cascade contributing at these distances therefore near-field ($<3.5$ km) effects and non-universal coherence loss from lateral spread are much less relevant. Secondly the causal structure of arrival times from different slices leads to substantial destructive interference where most of the shower contributes purely constructively on the Cherenkov ring. This destructive interference adds many opportunities for local errors to cancel each other out as our rescaling does not adjust the pulse shape.

With the exception of these large lateral distances a simple rescaling does not yield the desired synthesis accuracy, in particular because the errors represent a systematic deviation in our model or the used templates rather than physically real shower-to-shower fluctuations.

\section{Shower Evolution Corrections}

By comparing results for several template showers it becomes readily apparent that the synthesis quality decreases with increasing difference in $X_\mathrm{max}$. Thus our refinements focus on accounting for differences in cascade evolution between different showers. Our previous proposal of matching slices by shower evolution \cite{2017ICRC...35..307B} ultimately falls short of expectations due to the shortcomings of comparing slices of varying geometric size versus column depth and the increased risk of encountering unequal near-field scaling.

The simplest solution for equal geometries is to directly compare the evolution of different showers and to derive a scalar correction. For example one can observe a reasonable correlation between the peak amplitude or energy within each slice divided by the particle number and the depth of shower maximum or the closely related relative evolution stage and shower age. While this necessarily leads to \textit{some} improvement a comparison of individual slice time series reveals a change in pulse shape which also causes a scalar correction to fail at the previously well-matched large lateral distances due to destructive interference.

While the sum of slice time series (and much of the underlying CORSIKA/CoREAS) operates in the time domain deformations of narrow pulses are far more convenient to handle in the frequency domain. The amplitude spectrum of a single slice time series can vary greatly in shape however there will always be substantial noise caused by a combination of numerical thinning and the physical reality of observing a finite number of particles. At high frequencies only this incoherent noise remains while the coherent emission falls off exponentially, however the cutoff point depends on the propagation geometry. While we have ensured to fulfill the Nyquist sampling criterion for our desired operating bandwith of 0-500 MHz to avoid further numerical errors the spectral cutoff can in fact fall inside this band for numerous slices and therefore must be taken into consideration. Based on known models for physically observable pulse spectra \cite{2005APh....24..116H, 2019arXiv190511185W} we choose to fit our slice amplitude spectra using the function
\begin{equation}
\label{eq:ampfit}
A_\mathrm{slice}(f) = A_0 \cdot \exp( b \cdot f + c \cdot f^2 ) + d
\end{equation}
with
\begin{equation}
\sqrt{d} = \max \left[10^{-9} \cdot \left(\frac{X}{400 \mathrm{g}/\mathrm{cm}^2} - 1.5\right) \cdot \exp\left(1 - \frac{r}{40000 \mathrm{cm}}\right), 0 \right]
\end{equation}
tuned to represent the noise floor for vertical proton primaries of $10^{17}$eV. It should be noted we fit the electric field divided by the particle number, therefore the quantitative values of $A_0$ and $d$ do not match the expected physical scale and the first-order scaling with the cosmic ray primary energy already is accounted for. The noise only becomes relevant in the later stages of the cascade when secondary particles have had a significant amount of time to drift apart and lose coherence coupled with their physical proximity to the antenna yielding higher absolute signal strengths. Therefore our model ramps down and then entirely removes the noise term for earlier slices.
\begin{figure}
\begin{center}
\includegraphics[width=0.45\textwidth]{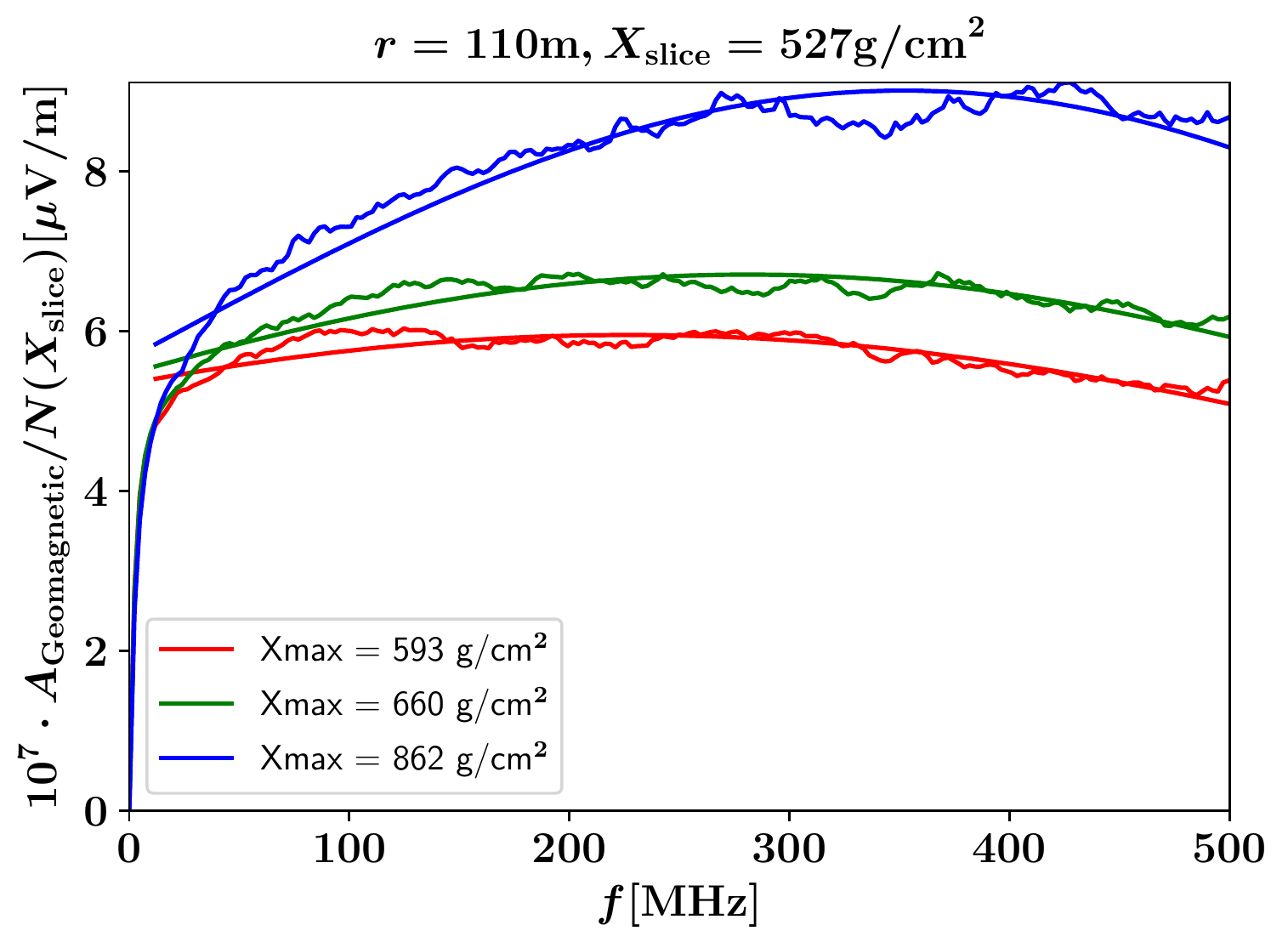}
\includegraphics[width=0.45\textwidth]{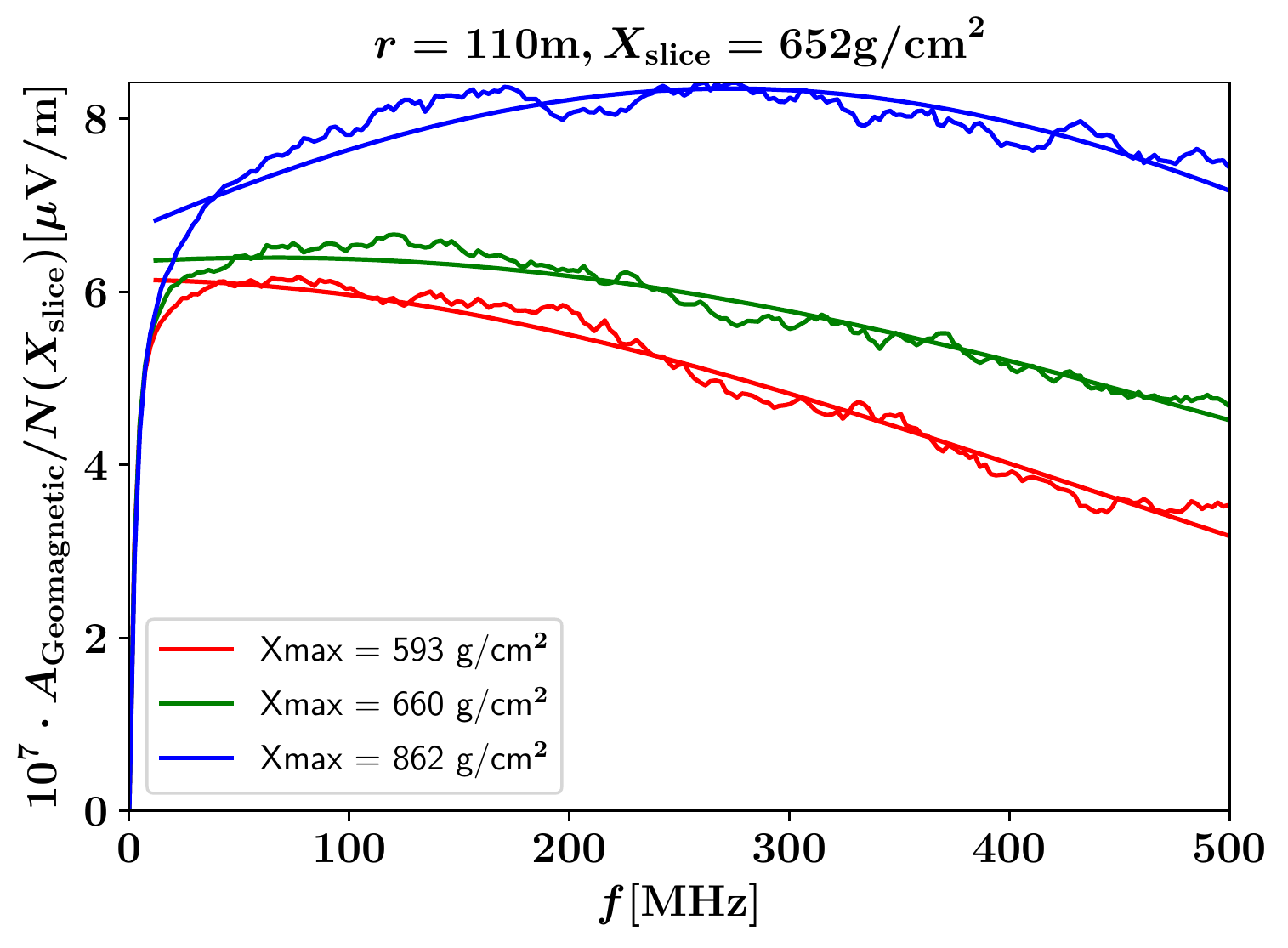}
\includegraphics[width=0.45\textwidth]{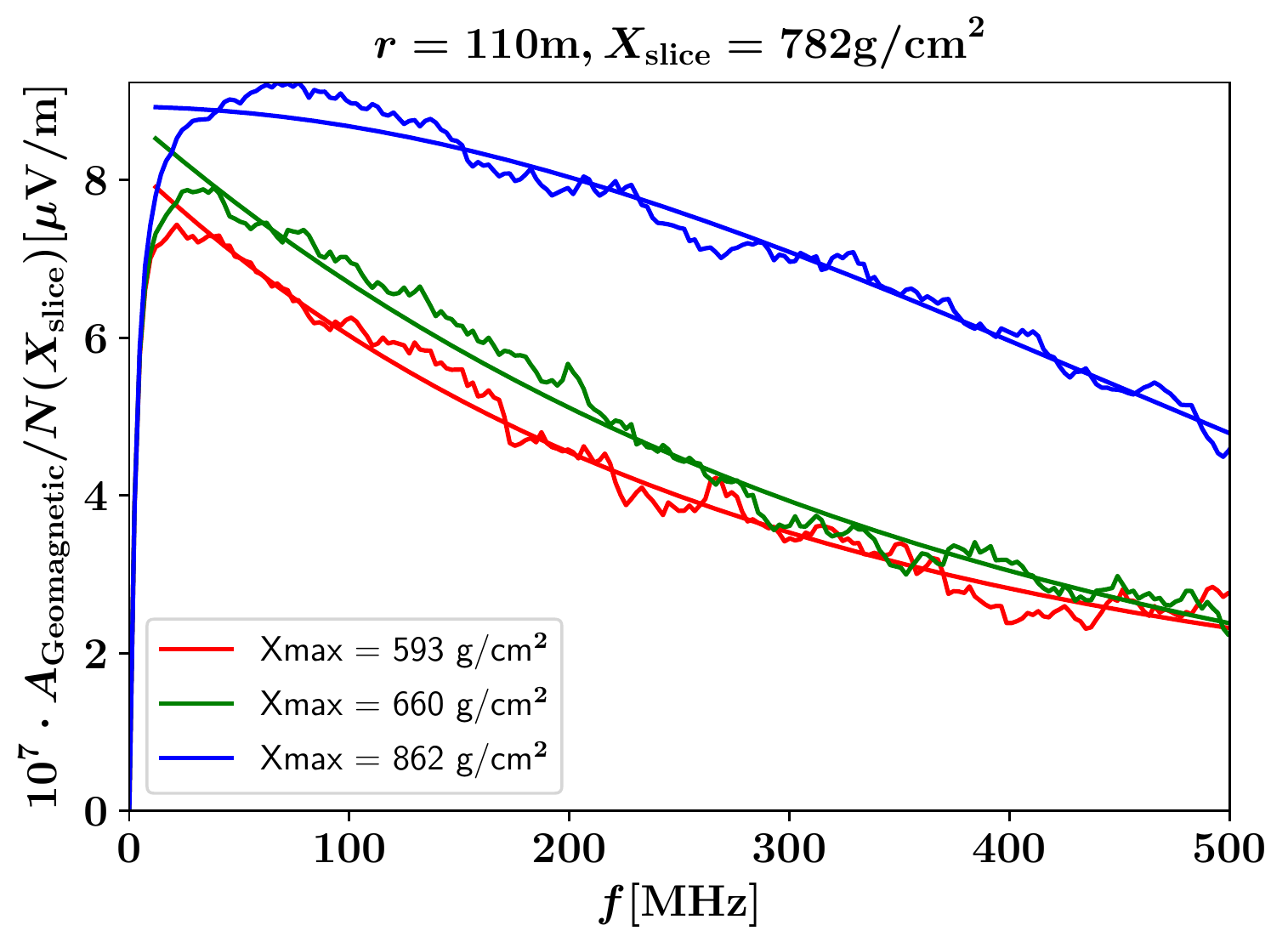}
\includegraphics[width=0.45\textwidth]{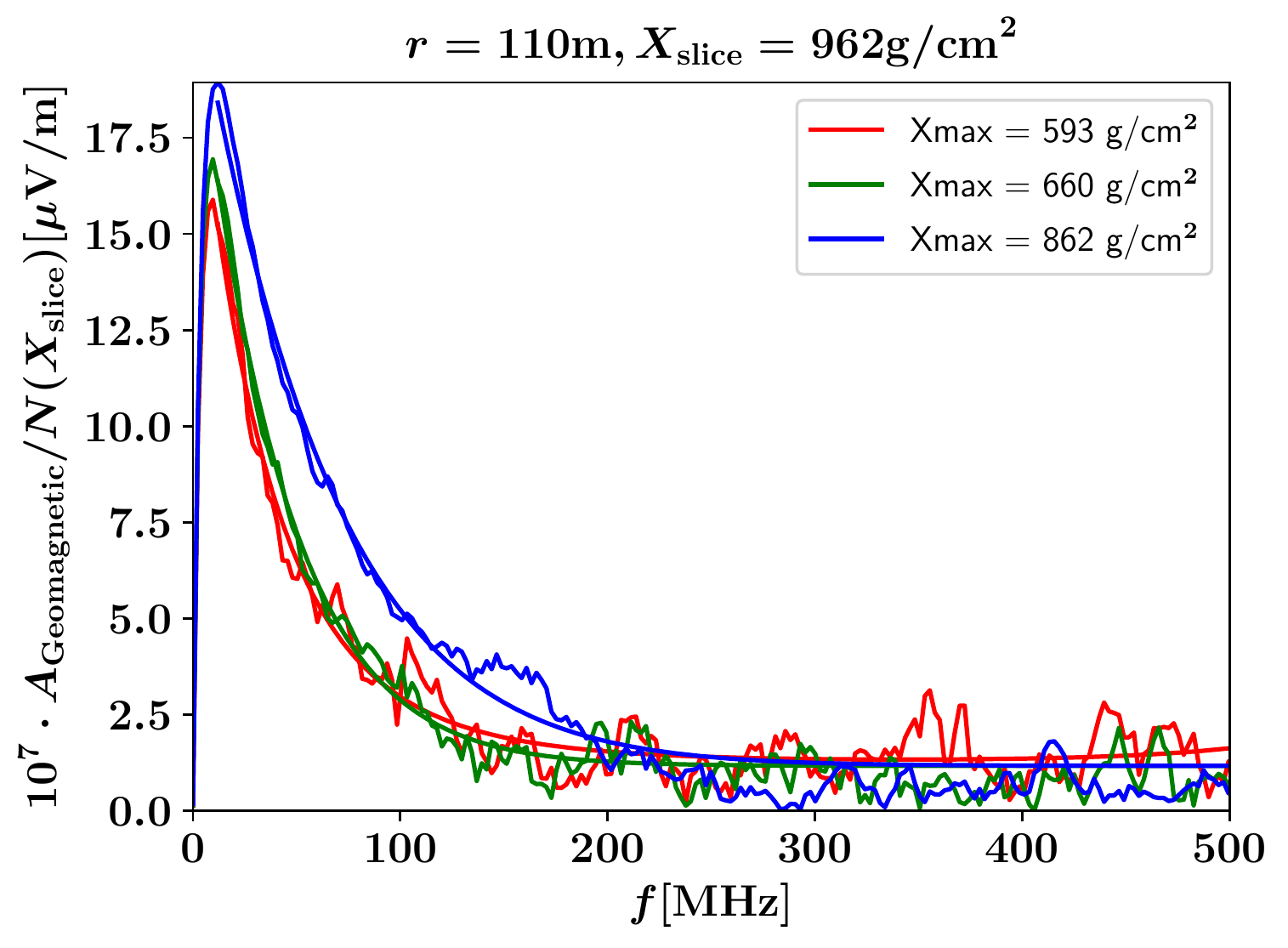}
\caption{Fits of amplitude spectra for different slices and different shower maxima but all from the same antenna located on the Cherenkov ring. The hierarchical ordering by $X_\mathrm{max}$ is consistent for other lateral distances as well.}
\label{fig:ampspec}
\end{center}
\end{figure}

We demand both $A_0$ and $d$ be strictly nonnegative but allow $b$ and $c$ to change signs in order to capture spectra of different shapes, fitting the spectral coefficients $A_0$, $b$ and $c$ independently for all slices and sample antenna positions in each simulation. This works well in our chosen frequency range of 0-500 MHz but care needs to be taken at larger viewing angles or very early or late evolution stages. The early stages are irrelevant as the very low number of particles leads to insignificant amplitudes but for the very late evolution at $X>950$ g/cm$^2$ we encounter numerical divergences as the incoherent noise begins to dominate the fit. 

For our refined synthesis model we compare the spectral coefficients $A_0$, $b$ and $c$ for many showers to their respective shower maximum $X_\mathrm{max}$ and derive an analytic fit for each slice and antenna position separately. Thus we obtain our amplitude spectrum model $A(X, \vec r, X_\mathrm{max}, f)$ from Eq. \ref{eq:ampfit} using $A_0(X, \vec r, X_\mathrm{max}),~ b(X, \vec r, X_\mathrm{max}),~ c(X, \vec r, X_\mathrm{max})$ where the $X_\mathrm{max}$ dependence can be evaluated analytically to cover the full continuum of potential inputs. We also observe this relation to be universal within shower-to-shower-fluctuations when comparing proton and iron primaries between $10^{17}$ and $10^{19}$ eV. Pending detailed verification with less numerical thnning in the simulations this implies the observable differences in the radio output due to primary species and energy result entirely from the development of the longitudinal profile $N(X)$.

\section{Refined Template Synthesis}

\begin{figure}
\begin{center}
\includegraphics[width=0.4\textwidth]{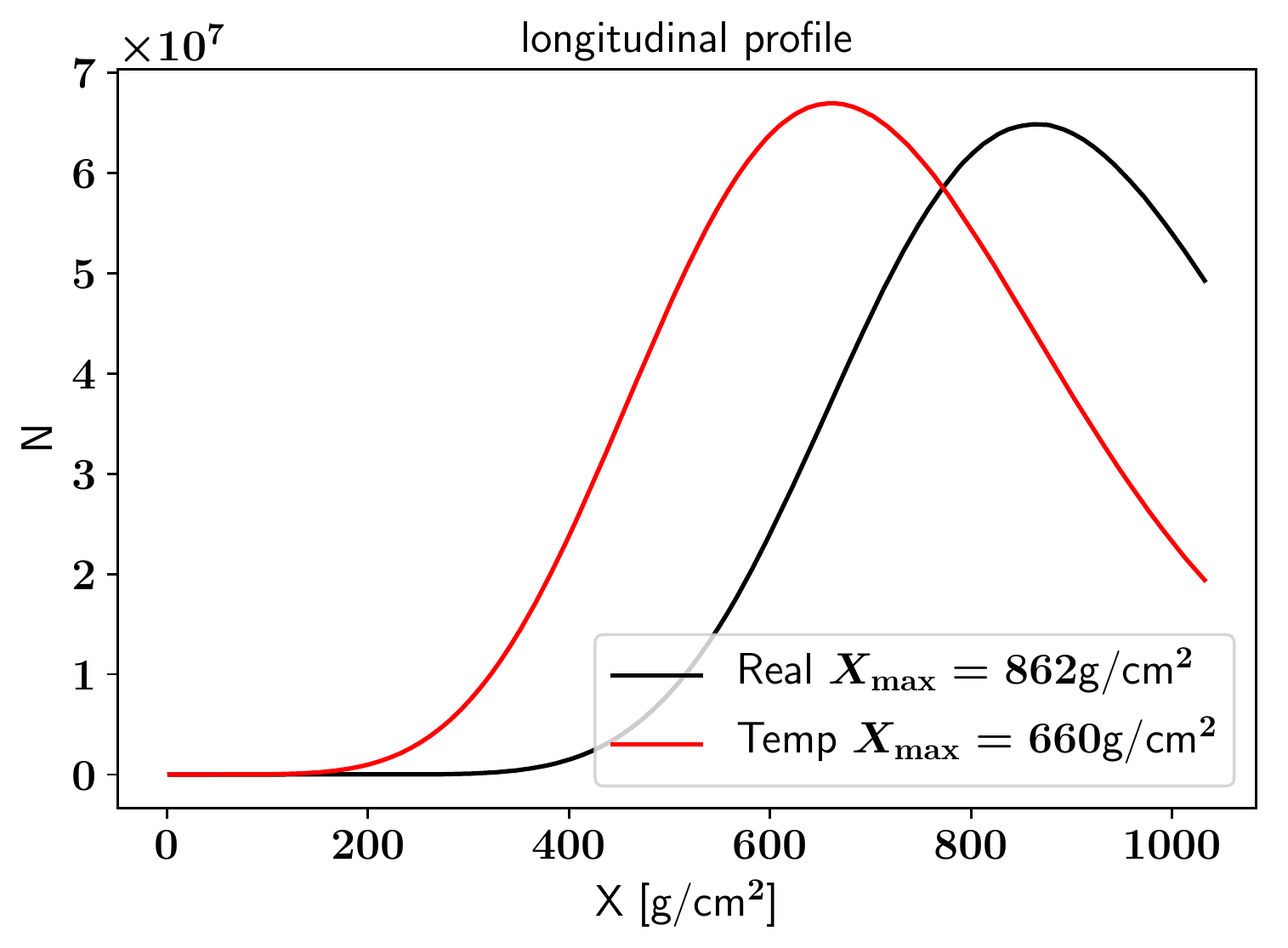}
\includegraphics[width=0.45\textwidth]{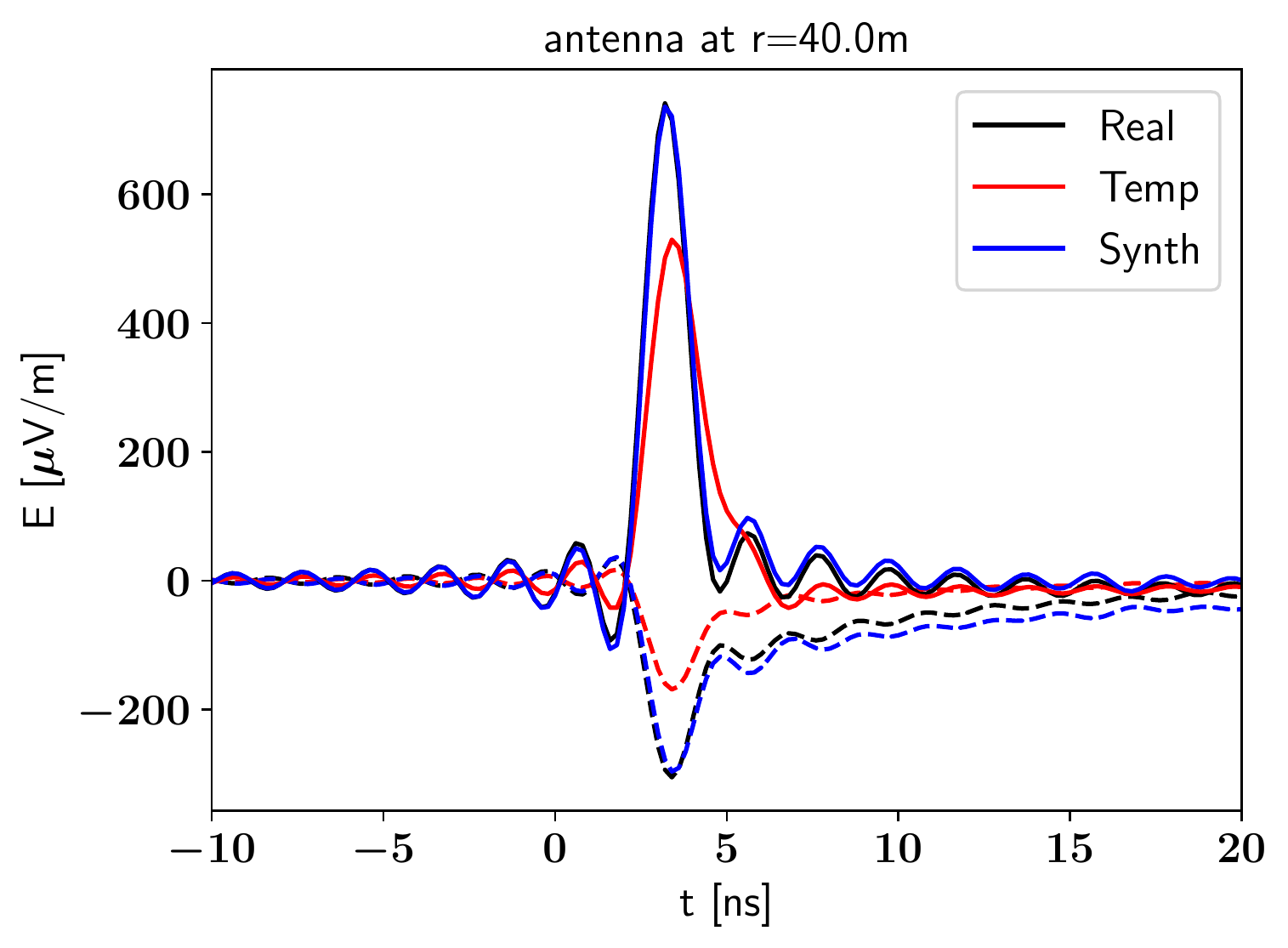}
\includegraphics[width=0.45\textwidth]{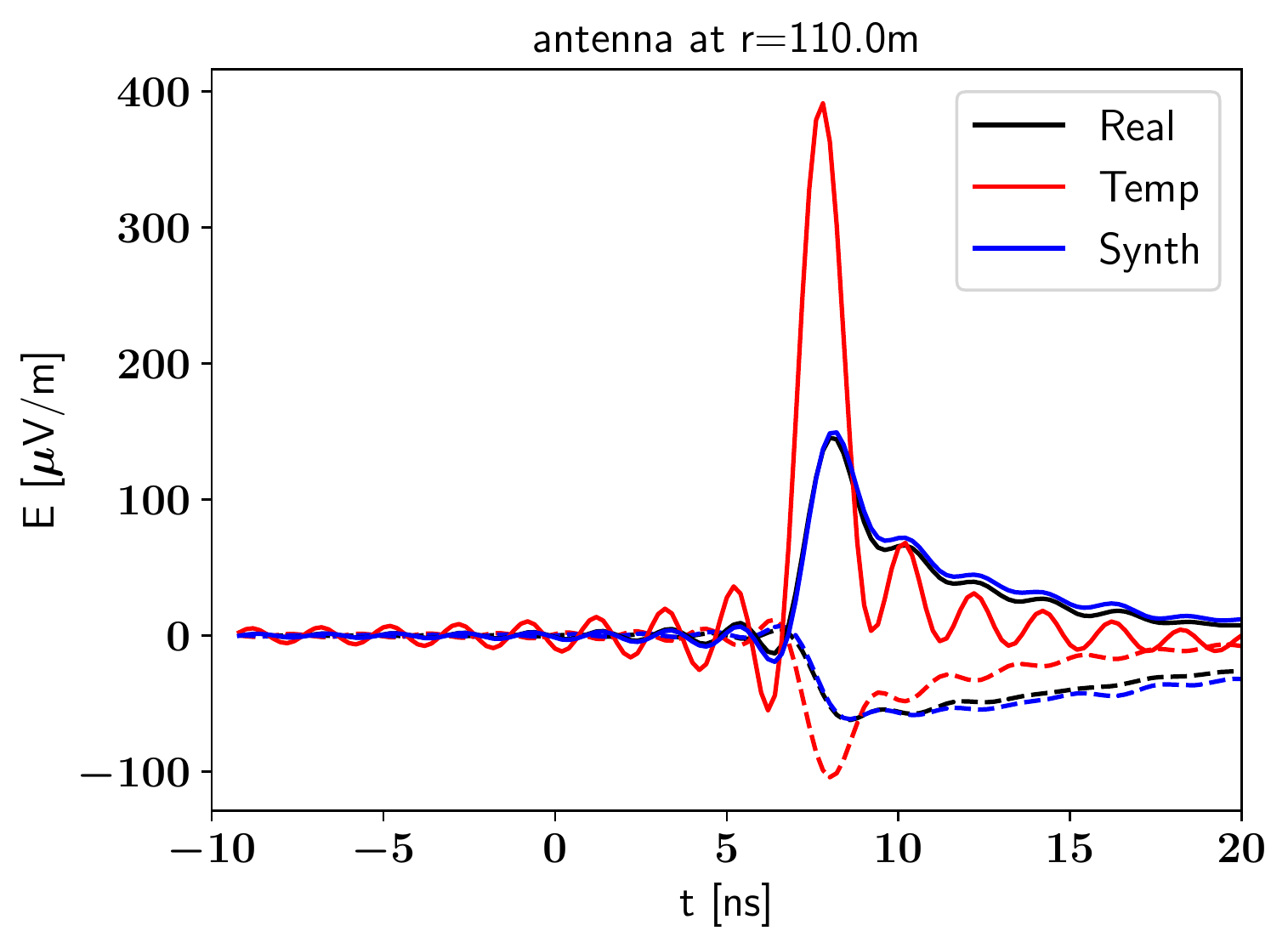}
\includegraphics[width=0.45\textwidth]{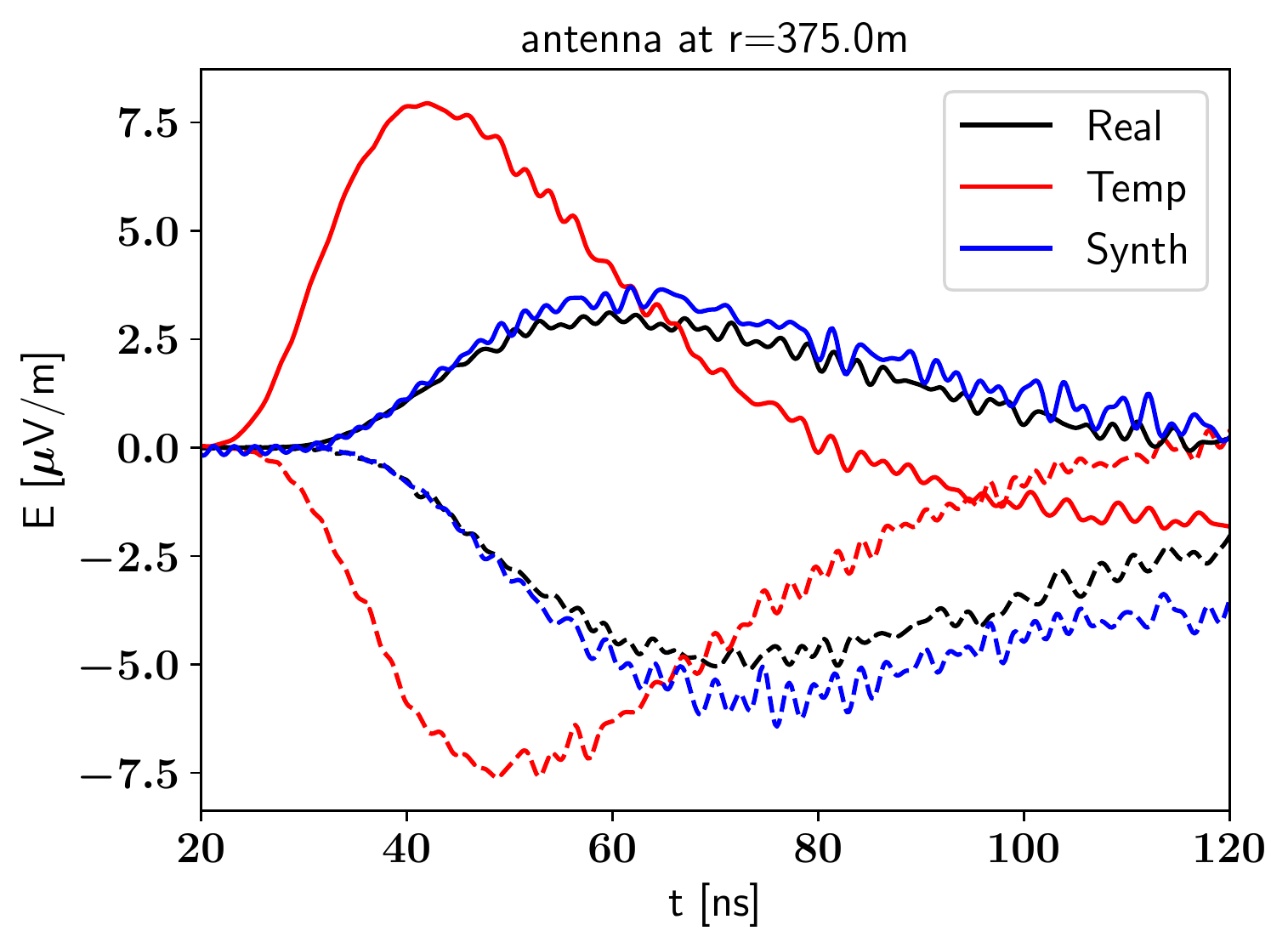}
\caption{Advanced synthesis with fitted amplitude spectrum correction following the same layout as Fig. \ref{fig:ssynth}. Both polarisation components match almost perfectly, although at 375m the independent fits for different slices appear to be destroying longitudinal coherence as this synthesis result is less accurate than the simple approach.}
\label{fig:asynth}
\end{center}
\end{figure}

We now use the fitted amplitude spectra in our additional frequency-dependent linear rescaling factor according to Eq. \ref{eq:asynth}. This works very well even for large differences in $X_\mathrm{max}$ as shown in Fig. \ref{fig:asynth} and represents the most accurate configuration of our current model.
\begin{equation}
\label{eq:asynth}
\vec{E}^\mathrm{Synth} (\vec r, t) = \sum_X \frac{N^\mathrm{Real}(X)}{N^\mathrm{Temp}(X)} \cdot \mathcal{F}^{-1} \left[ \frac{A(X, \vec r, X^\mathrm{Real}_\mathrm{max}, f)}{A(X, \vec r, X^\mathrm{Temp}_\mathrm{max}, f)} \cdot {\vec E}^\mathrm{Temp}_\mathrm{slice}(X,\vec r,f) \right]
\end{equation}

It should be noted the rescaling can encounter numerical issues when using analytical fitted functions which converge to zero too quickly, therefore the current procedure is not yet suitable for automatic bulk application. The loss of near-perfect accuracy at large lateral distances results from contributions below 20 MHz which are not observable in practice. Where synthesis accuracy including very low frequencies is the only matter of importance we could return to the simple rescaling approach for large lateral distances.

On the other hand if our rescaling were perfect dividing the complex electric field in the frequency domain by both the particle number and the amplitude spectrum fit should leave only a phase factor and incoherent noise. To first order the phase spectrum only contains a linear gradient representing the arrival time of the pulse which can also be predicted analytically. The achieved accuracy of our synthesis model suggests it might be possible to construct an entirely analytical model from empirically fitted coefficients and the residual incoherent signals could be reprocessed into an uncertainty estimate. 

\section{Conclusion}

We have developed a fast and precise radio signal synthesis model within our controlled test conditions. The next step is to generalise the core principles to a parameter space viable for practical application. Our current amplitude model is entirely empirical thus we would like to relate it to physical properties of the air shower cascade.

If the demonstrated synthesis accuracy can be maintained under realistic conditions our model could be applied as a replacement for computationally expensive MC simulations in most common analysis methods. In particular it would become feasible to apply minimum-$\chi^2$ shower maximum reconstruction \cite{2014PhRvD..90h2003B} or similar sampling-based techniques at higher primary energies and for datasets measured by large numbers of antennas.

\printbibliography

\end{document}